\documentclass[twocolumn,showpacs,prl]{revtex4}
\usepackage[pdftex]{graphicx}
\usepackage{natbib}
\begin{document}
\title{Tilted potential induced coupling of localized states in a graphene nanoconstriction}

\author{M. R. Connolly$^1$, K. L. Chiu$^1$, A. Lombardo$^{2}$, A. Fasoli$^2$, A. C. Ferrari$^2$, D. Anderson$^1$, G. A. C. Jones$^1$, C. G. Smith$^1$}

\affiliation{$^1$Cavendish Laboratory, Department of Physics, University of Cambridge, Cambridge, CB3 0HE, UK}
\affiliation{$^2$Department of Engineering, University of Cambridge, Cambridge, CB3 OFA, UK}

\date{\today}

\begin{abstract}

We use the charged tip of a low temperature scanning probe microscope to perturb the transport through a graphene nanoconstriction. Maps of the conductance as a function of tip position display concentric halos, and by following the expansion of the halos with back-gate voltage we are able to identify an elongated domain over the nanoconstriction where they originate. Amplitude modulations of the transmission resonances are correlated with the gradient of the tip-induced potential and we analyze this in terms of modified coupling between localized states.

\end{abstract}

\maketitle

The demand for materials capable of realising the next generation of electronic and photonic devices continues to fuel interest in the electronic properties of graphene \citep{Geim2007, Bonaccorso2010}. The single-electron control offered by quantum dots makes them ideal for examining properties of Dirac quasiparticles such as the g-factor \citep{Guttinger2008}, excitation spectra \citep{Molitor2010}, and spin relaxation times \citep{Buitelaar2008}. Graphene dots typically consist of an etched island which is tunnel coupled to large area leads by ultranarrow ($<$30 nm) constrictions \citep{Ponomarenko2008,Stampfer2008}. Broad and irregular modulations of the coulomb blockade oscillations obtained from such structures have been attributed to disorder-induced dots within the constrictions themselves \citep{Stampfer2008}. Although their size, spacing, and precise origin is currently unclear, at sub-30 nm lengthscales it is likely that potential inhomogeneities and edge roughness play a role in creating localized states \citep{Schnez2010, Stampfer2009, Todd2009}. The random transport properties associated with such unintentional localization is not only problematic for analyzing quantum dot structures but also presents challenges for the development of nanoribbon electronics \citep{Chen2007}. In contrast to large area graphene sheets whose on-off conductance ratio is too low for applications such as switches and transistors in logic applications, narrow channels do exhibit a length- and width-dependent transport gap where the conductance is suppressed for a range of source-drain bias around the charge neutrality point \citep{Han2007, Todd2009, Molitor2009}. This gapped region is larger than estimates based on confinement effects alone, however, and also exhibits sharp resonances at low temperature \citep{Molitor2009,Todd2009, Liu2009}. Hence the focus has recently shifted away from tailoring transport properties using edge lattice symmetries and more towards understanding the role of disorder \citep{Han2007, Han2010}. Transport through narrow graphene channels in the presence of disorder has been explored within different theoretical and semi-empirical frameworks, including  Anderson localisation \citep{Evaldsson2008}, percolation models \citep{Adam2008}, disorder and edge-roughness induced quantum dot formation \citep{Sols2007,Han2010,Todd2009, Molitor2009}, and lateral confinement \citep{Han2007}. Since these mechanisms are partly distinguished by \textit{where} localization occurs, local probes offer a powerful way to discriminate between them. 
\begin{figure}	
\includegraphics{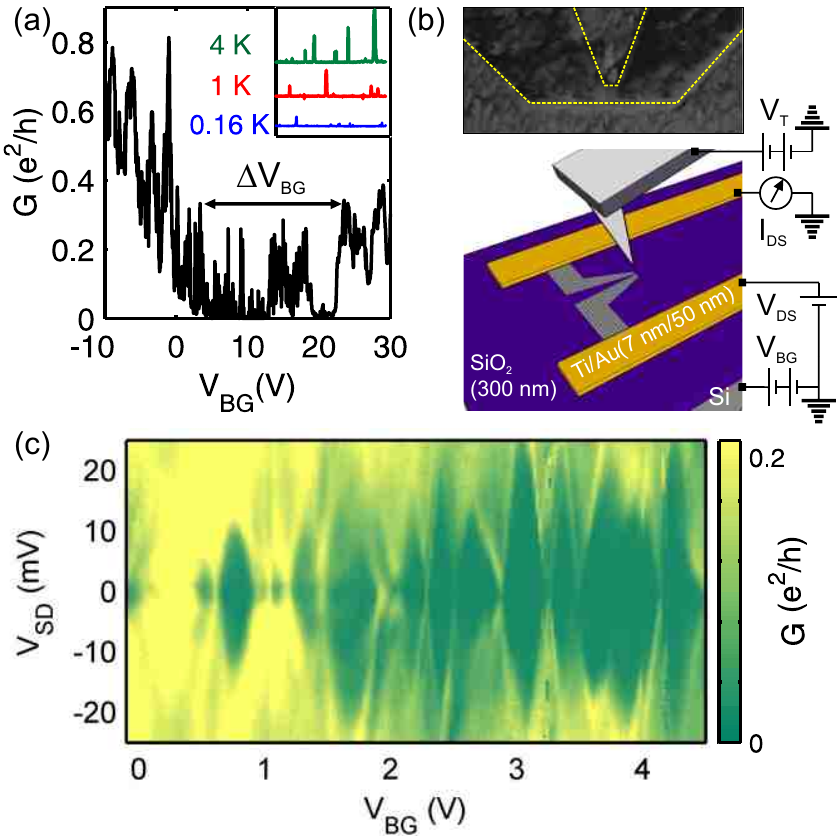}
\caption{(a) Conductance of the nanoconstriction as a function of $V_{BG}$ at $T$=8 K ($V_{SD}$=5 mV). Inset: plots of the zero-bias conductance (excitation voltage 300 $\mu V$) within the transport gap ($V_{BG}$= 7-10 V) at different temperatures. (b) Setup used to perform scanning gate microscopy. Inset: Atomic force micrograph ($\approx$ 1 $\times$ 0.5 $\mu$m$^2$) showing the 90 nm wide constriction. (c) Conductance as a function of $V_{SD}$ and $V_{BG}$. Diamonds of suppressed conductance at low $V_{SD}$ indicate the presence of coulomb blockaded islands with characteristic charging energy of $\approx$15-20 meV.}  
\label{Fig:Fig1}
\end{figure}

In this work we use the charged tip of a scanning probe microscope to perturb the transport through a lithographically defined graphene nanoconstriction. Although the broad lineshape of the resonances hampers us from resolving their exact location, we are nonetheless able to detect the presence of multiple dots within the channel via the effect of the tilted potential from the tip. 

Our graphene flakes are mechanically exfoliated from natural graphite onto a highly doped Si substrate capped with a 300 nm thick SiO$_{2}$ layer. Optical microscopy \citep{CasiraghiNL} and Raman spectroscopy \citep{Ferrari2006} are used to locate and confirm the flakes are monolayers. Two 10/50 nm thick Ti/Au contacts were patterned using lift-off processing and a $\approx$90 nm wide channel connected by tapered leads was etched using an O$_2$ plasma [Fig. \ref{Fig:Fig1}(b).] Figure \ref{Fig:Fig1}(a) shows the conductance $G$ of the device as a function of voltage $V_{BG}$ applied to the Si back-gate at $T$= 8 K. As expected \citep{Stampfer2008}, $G$ is strongly suppressed for a range $\Delta V_{BG}$ around the neutrality point ($V_{NP}$$\approx$12 V) except for irregularly spaced resonances where $G$ increases over a narrow range of $V_{BG}$. The density of resonances decreases at lower temperature while their lineshape remains roughly constant [inset, Fig. \ref{Fig:Fig1}(a).] At larger source-drain voltage ($V_{SD}$) the conductance between the resonances increases, leading to $\Delta V_{SD}$=15-20 mV high diamond shaped regions where current is blockaded [Fig. \ref{Fig:Fig1}(c).]

\begin{figure}
\includegraphics{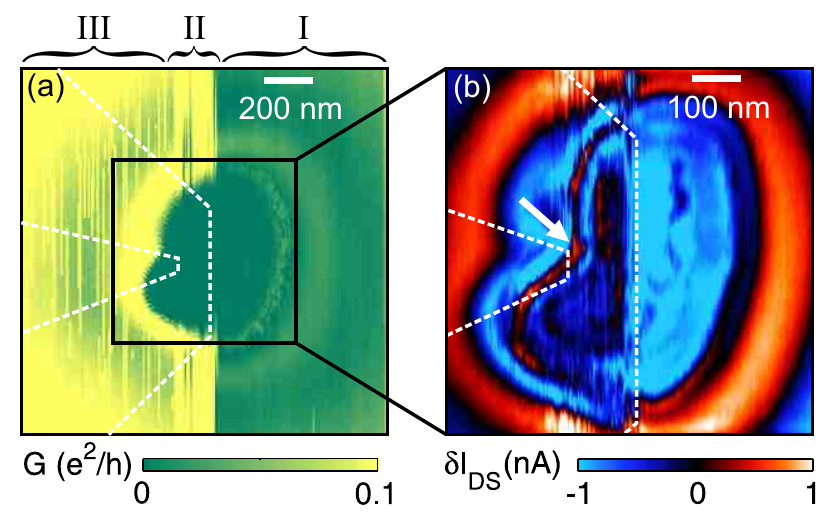}
\caption{(a) Scanning gate image of the nanoconstriction captured in tapping mode ($V_T$=0.5 V, $V_{BG}$=0 V, $V_{SD}$=5 mV.) (b) Transconductance scanning gate image captured with $V_T$=50 mV @ 700 Hz. The arrow indicates the position of the indent used to align lift-mode scanning gate images. White dashed outlines indicate the edge of the nanoconstriction extracted from the simultaneously captured topographic image.}     
\label{Fig:Fig2}
\end{figure}

A schematic of our setup used to perform scanning gate microscopy (SGM) is shown in Fig. \ref{Fig:Fig1}(b). We use a Pt/Ir coated cantilever (NanoWorld ARROW-NCPt) with a nominal tip radius of 15 nm. To perform SGM a current is driven through the device and its conductance is recorded as a function of tip position. Figure \ref{Fig:Fig2}(a) shows a typical SGM image \citep{Horcas2007} captured over the constriction in tapping mode with $V_T$=0.5 V and $V_{BG}$=0 V. Elongated halos alternating between enhanced and suppressed $G$ encircle the constriction. Each halo corresponds to the locus of points where the tip's contribution to the electrostatic potential is sufficient to overcome the coulomb blockade \citep{Woodside2002}. The halos are continuous in region \textit{I} but are broken by a stripe in region \textit{II} where the tip crosses the outer graphene edge on either side. This stripe rotated with the scan direction and was always parallel to the fast scan axis. A region with these properties was also observed in SGM of subsurface quantum dots \citep{Pioda2004}, where it was attributed to tip-induced rearrangements of charge. Although their elongated axis is still aligned with the channel, in region \textit{III} the halos are indented where the inside edge of the constriction meets the tapered lead. This indent is highlighted in Fig. \ref{Fig:Fig2}(b), which shows an SGM image of the transconductance captured under the same conditions as Fig. \ref{Fig:Fig2}(a), but with a modulated tip bias to enhance the contrast \citep{Connolly2010}. There is good agreement between the origin of the halos and the topographic location of the constriction, presumably due to the proximity of the tip in tapping mode  \citep{Schnez2010}. While this mode was useful for relating SGM structures to the underlying topography, it was unsuitable for continuous use as the distribution of resonances in $G(V_{BG})$ changed between successive scans. To avoid this instability we performed all subsequent scans in ``lift-mode'' with the static tip $\approx$30 nm from the surface, and used the indent [arrow, Fig. \ref{Fig:Fig2}(b)] as a reference point for overlaying the topographic outline. 

\begin{figure}
\includegraphics{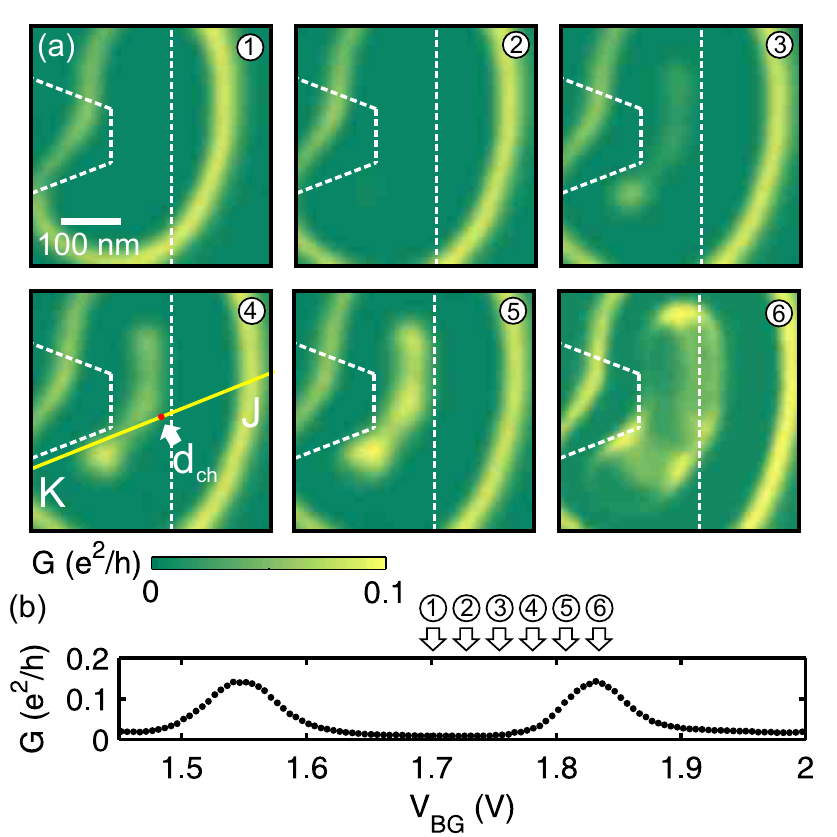}
\caption{(a) Sequence of lift-mode scanning gate images showing the evolution of the conductance halos with back-gate voltage ($V_T$=2 V). Dashed outline indicates the approximate location of the nanoconstriction. (b) Linear conductance as a function of back-gate voltage. Numbered arrows indicate the back-gate voltages at which the corresponding images in (a) were captured.}     
\label{Fig:Fig3}
\end{figure}

To establish where the conductance haloes originate we park the biased tip ($V_T$= 2 V) over the channel and tune $V_{BG}$ such that $G$ lies between two resonances [Fig. \ref{Fig:Fig3}(b).] The first resonance manifests as the outer halo in image 1 of Fig. \ref{Fig:Fig3}(a). With increasing $V_{BG}$ the second resonance emerges as a disordered domain, which is aligned with the channel and extends into the leads with weak and variable contrast [images 2-5, Fig. \ref{Fig:Fig3}(a)]. Once $G$ is tuned to the peak of the second resonance the domain is roughly the same width as the channel and splits into a single halo [image 6, Fig. \ref{Fig:Fig3}(a).] This observation is in good agreement with recent scanning gate work on similar graphene nanostructures \citep{Schnez2010}. The central position of the domain and sensitivity to tapping-mode-induced changes in the electrostatic environment suggest that potential inhomogeneity-, rather than edge roughness-, induced localization is the dominant mechanism in our structure \citep{Ni2009,Todd2009}. While the shape of the domain depends on the profile of the tip potential as well as the shape of the region of localized states \citep{Kicin2005}, we assume that tip-induced distortions play less of a role here because the orientation, width, and location of the domain compare favourably with the constriction itself. 

\begin{figure}
\includegraphics{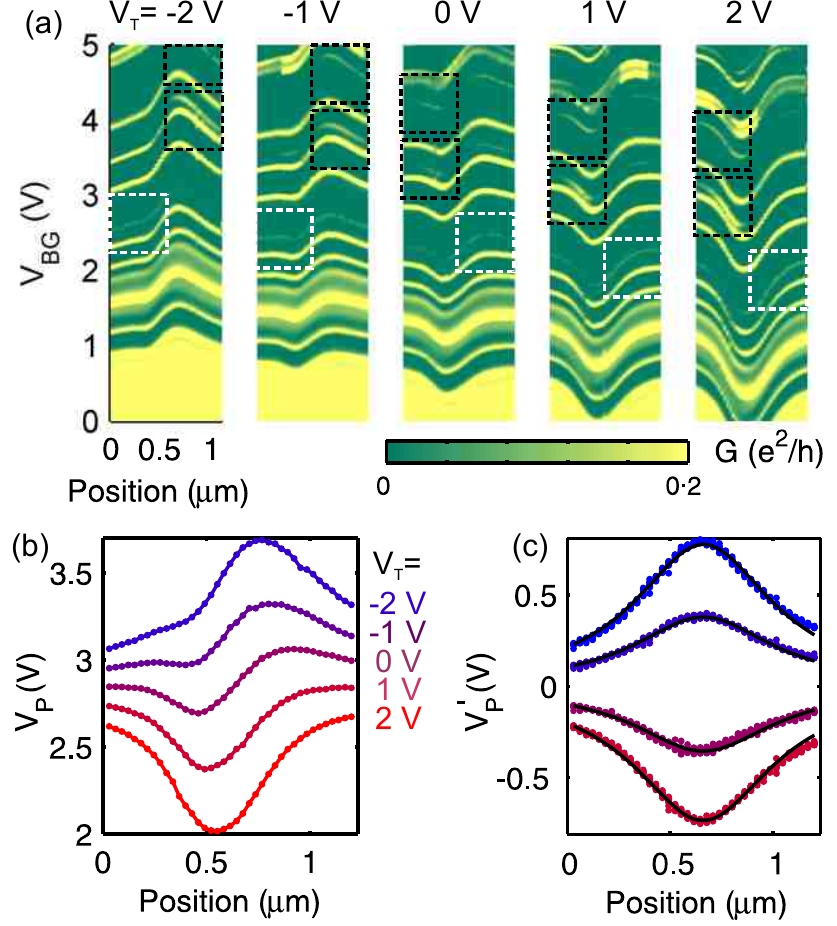}
\caption{(a) Linear conductance as a function of back-gate voltage and tip position along the line $JK$ in Fig. \ref{Fig:Fig3} at different tip voltages. Boxes with dashed outlines highlight resonances which exchange side upon reversing the sign of the tip potential. (b) Plots of the back-gate voltage of a typical resonance as a function of position along $JK$ at different tip voltages. (c) The same plot as (b) for nine of the resonances in (a) with the zero-$V_T$ variation subtracted. Solid lines are fitted Lorentzians.}    
\label{Fig:Fig4}
\end{figure}

While the overlapping coulomb diamonds and a reduction in the density of resonances at lower temperature implies that localization is due to chargeable islands separated by tunnel barriers \citep{Dorn2004}, our SGM images do not exhibit the characteristic signatures of multiple quantum dots, such as interlocking or anti-crossing halos \citep{Schnez2010, Bieszynski2007, Woodside2002}. These observations are reconciled in the case of islands with insufficient capacitive coupling to create charging effects between them \citep{Schnez2010}. The latter may reflect the elevated tunnel coupling ($\Gamma$) of the barriers in this region of $V_{BG}$, an interpretation supported by the $T$ independence of the peak lineshape, which suggests $\hbar\Gamma > k_B T$. Furthermore, one would expect the effect of a broad tip potential to be similar to a global back-gate if the islands are closely spaced, leading to concentric, rather than interlocking, halos around the domain seen in Fig. \ref{Fig:Fig3}(a). 

\begin{figure}
\includegraphics{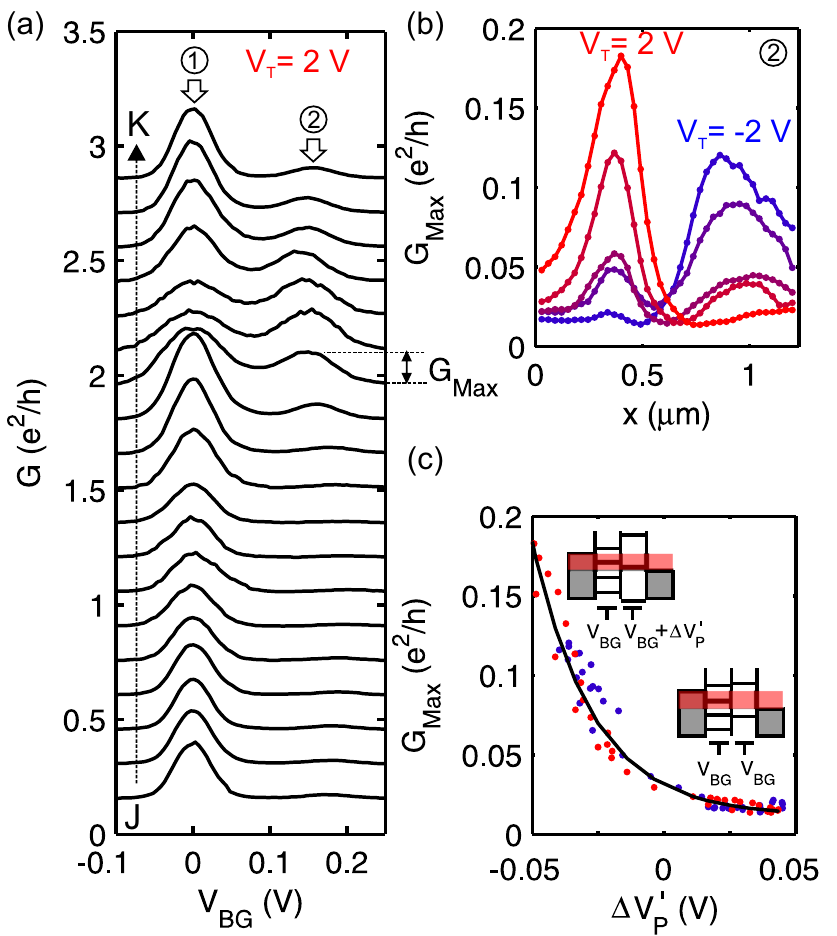}
\caption{(a) Conductance as a function of back-gate voltage with the tip at different positions along $JK$ in Fig. \ref{Fig:Fig3}(a) with $V_T$=2 V. Each curve has been vertically offset for clarity and horizontally offset such that peak (1) is at $V_{BG}$= 0 V. (b) Maximum conductance of resonance (2) as a function of tip position for $V_T$= 2 V (red) and -2 V (blue). (c) Maximum conductance of resonance (2) as a function of the gradient of the corrected tip-induced potential. Inset: schematic depictions showing the offset in the ladder of states of two dots in series with different coupling to the back-gate.}    
\label{Fig:Fig5}
\end{figure}

In the absence of pronounced multiple-dot features in the shape of the halos we focus on the tip-position dependent amplitude modulations within the domain of Fig. \ref{Fig:Fig3}(a). The tip was moved in $30$ nm steps along the line $JK$ in Fig. \ref{Fig:Fig3}(a) and the back-gate voltage was swept at each point. The resulting $G(V_{BG},x)$ maps are plotted in Fig. \ref{Fig:Fig4}(a) at different tip voltages. Fig. \ref{Fig:Fig4}(b) shows a typical set of $V_{P}(x)$ curves, where $V_{P}$ denotes the position of a resonance in back-gate voltage. By subtracting from each $V_{P}(x)$ its behaviour at $V_T$=0 V we obtain a set of corrected data $V_{P}^{'}(x)$ which are symmetrical and well fitted by Lorentzians given by $V_{P}^{'}=A/(4(x-d_{ch})^2+w^2)$, where $d_{ch}$ is the centre of the Lorentzian along $JK$, $w$ is its full-width half maximum, and $A$$\approx$$\beta V_{T}$, with $\beta$$\approx$0.38$\pm0.01$. We find $d_{ch}$$\approx$675$\pm$10 nm and $w$$\approx$805$\pm$20 nm for thirty six curves corresponding to nine of the resonances in Fig. \ref{Fig:Fig4}(a). The value of $w$ is in good agreement with the width of tip-induced potentials reported previously \citep{Gildemeister2007, Wilson2008}. The centre $d_{ch}$ has been indicated in Fig. \ref{Fig:Fig3}(a). We attribute the variation of $V_P$ at zero $V_T$ to a combination of screening, contact potential difference,  charged debris on the tip, and capacitive coupling between the tip and the back-gate \citep{Gildemeister2007, Kicin2005, Bieszynski2007}. Note that the symmetry of $V_{P}$ in Fig. \ref{Fig:Fig4}(a) improves in the range of $V_T$ and $V_{BG}$ where the images in Fig. \ref{Fig:Fig3}(a) were captured, supporting our previous assertion that zero-$V_T$ distortions are minimized.

The maximum conductance $G_{Max}$ of the resonances highlighted by black dashed boxes in Fig. \ref{Fig:Fig4}(a) is enhanced (suppressed) when the tip is closer to $J$ ($K$) for $V_T$=2 V, while the opposite behaviour is observed when $V_T$= -2 V. Line plots of $G(V_{BG})$ around the pair of resonances in the middle box are shown in Fig. \ref{Fig:Fig5}(a), with the tip-induced background variation in $V_{P}$ removed to facilitate comparison. Since exchanging both the side and sign of the tip potential preserves the in-plane electric field direction, i.e. $E(-x)=-E(x)$, the observed asymmetry can be described in terms of the tilted potential induced along $JK$. To confirm this, in Fig. \ref{Fig:Fig5}(b) we plot $G_{Max}$ as a function of $\Delta V_{P}^{'}$, the gradient of the corrected tip-induced potential shown in Fig. \ref{Fig:Fig4}(b), which is related to the in-plane field by $E=\Delta V_{P}^{'}/\Delta x$. The data follow the same dependence on $\Delta V_{P}^{'}$ for $V_T$= $\pm$2 V, confirming that in this range the tilted potential controls the amplitude of this resonance. The data is well described by an exponential function $G_{Max}\approx e^{-\Delta V_{P}^{'}/\gamma}+D$ where $D$ $\approx$0.01 $e^2/h$ accounts for the non-zero conductance floor and $\gamma$$\approx$ 0.02 V. Similar behaviour is observed for the resonance in the other black box in Fig. \ref{Fig:Fig4}(a) and the opposite for the resonance in the white dashed box. Such variations in $G_{Max}$ are usually ascribed to tip-induced modulations in the tunnel barriers, and are therefore compatible with single dot behaviour \citep{Fallahi2005}. The exponential dependence on $\Delta V_{P}^{'}$ is moreover highly suggestive of tunnel barrier suppression. However, in light of the multiple-dot behaviour in the transport data we propose an alternative scenario. The different charging energies and coupling between the back-gate and the dots leads to a random distribution of resonances in $G(V_{BG})$ as a dot with the weakest coupling to the back-gate modulates the amplitude of resonances from the more strongly coupled dot(s) \citep{Ruzin1992, Gallagher2010}. The tilted potential introduced by the tip effectively offsets the ladder of charge states in the dots relative to one another [inset, Fig. \ref{Fig:Fig5}(c).] As a result the alignment of states in different dots either increases [peak 2 in Fig. \ref{Fig:Fig5}(a)] or decreases [peak 1 in Fig. \ref{Fig:Fig5}(a), and resonance in the white dashed box in Fig. \ref{Fig:Fig4}(a)] depending on their initial detuning. Within this picture, $G_{P}(\Delta V_{P}^{'})$ traces the resonance of a weakly coupled dot.

To summarize, we have investigated the response of a graphene nanoconstriction to the local electrostatic potential of a scanning probe tip. Weakly invasive scanning affects the resonance distribution within the transport gap, suggesting that localization in graphene nanostructures is sensitive to the local electrostatic environment. The expansion of a single elongated conductance halo around the constriction implies that the localized states consist of a quantum dot induced by potential inhomogeneities within the channel. Transport data combined with measurements at different tip potentials allow us to identify multiple-dot behaviour by analyzing the effect of the tilted potential on the coupling between localized states. 

This work was financially supported by the European GRAND project (ICT/FET). ACF acknowledges funding from the European Research Grant NANOPOTS and the Royal Society.


\begin{thebibliography}{33}
\expandafter\ifx\csname natexlab\endcsname\relax\def\natexlab#1{#1}\fi
\expandafter\ifx\csname bibnamefont\endcsname\relax
  \def\bibnamefont#1{#1}\fi
\expandafter\ifx\csname bibfnamefont\endcsname\relax
  \def\bibfnamefont#1{#1}\fi
\expandafter\ifx\csname citenamefont\endcsname\relax
  \def\citenamefont#1{#1}\fi
\expandafter\ifx\csname url\endcsname\relax
  \def\url#1{\texttt{#1}}\fi
\expandafter\ifx\csname urlprefix\endcsname\relax\def\urlprefix{URL }\fi
\providecommand{\bibinfo}[2]{#2}
\providecommand{\eprint}[2][]{\url{#2}}

\bibitem[{\citenamefont{Geim and Novoselov}(2007)}]{Geim2007}
\bibinfo{author}{\bibfnamefont{A.~K.} \bibnamefont{Geim}} \bibnamefont{and}
  \bibinfo{author}{\bibfnamefont{K.~S.} \bibnamefont{Novoselov}},
  \bibinfo{journal}{Nature Materials} \textbf{\bibinfo{volume}{6}},
  \bibinfo{pages}{183} (\bibinfo{year}{2007}).

\bibitem[{\citenamefont{Bonaccorso et~al.}(2010)\citenamefont{Bonaccorso, Sun,
  Hasan, and Ferrari}}]{Bonaccorso2010}
\bibinfo{author}{\bibfnamefont{F.}~\bibnamefont{Bonaccorso}},
  \bibinfo{author}{\bibfnamefont{Z.}~\bibnamefont{Sun}},
  \bibinfo{author}{\bibfnamefont{T.}~\bibnamefont{Hasan}}, \bibnamefont{and}
  \bibinfo{author}{\bibfnamefont{A.~C.} \bibnamefont{Ferrari}},
  \bibinfo{journal}{arXiv:1006.4854v1}  (\bibinfo{year}{2010}).

\bibitem[{\citenamefont{Guettinger et~al.}(2010)\citenamefont{Guettinger, Frey,
  Stampfer, Ihn, and Ensslin}}]{Guttinger2008}
\bibinfo{author}{\bibfnamefont{J.}~\bibnamefont{Guettinger}},
  \bibinfo{author}{\bibfnamefont{T.}~\bibnamefont{Frey}},
  \bibinfo{author}{\bibfnamefont{C.}~\bibnamefont{Stampfer}},
  \bibinfo{author}{\bibfnamefont{T.}~\bibnamefont{Ihn}}, \bibnamefont{and}
  \bibinfo{author}{\bibfnamefont{K.}~\bibnamefont{Ensslin}},
  \bibinfo{journal}{arXiv:1002.3771v1}  (\bibinfo{year}{2010}).

\bibitem[{\citenamefont{Molitor et~al.}(2010)\citenamefont{Molitor, Knowles,
  Droescher, Gasser, Choi, Roulleau, Guettinger, Jacobsen, Stampfer, Ensslin
  et~al.}}]{Molitor2010}
\bibinfo{author}{\bibfnamefont{F.}~\bibnamefont{Molitor}},
  \bibinfo{author}{\bibfnamefont{H.}~\bibnamefont{Knowles}},
  \bibinfo{author}{\bibfnamefont{S.}~\bibnamefont{Droescher}},
  \bibinfo{author}{\bibfnamefont{U.}~\bibnamefont{Gasser}},
  \bibinfo{author}{\bibfnamefont{T.}~\bibnamefont{Choi}},
  \bibinfo{author}{\bibfnamefont{P.}~\bibnamefont{Roulleau}},
  \bibinfo{author}{\bibfnamefont{J.}~\bibnamefont{Guettinger}},
  \bibinfo{author}{\bibfnamefont{A.}~\bibnamefont{Jacobsen}},
  \bibinfo{author}{\bibfnamefont{C.}~\bibnamefont{Stampfer}},
  \bibinfo{author}{\bibfnamefont{K.}~\bibnamefont{Ensslin}},
  \bibnamefont{et~al.}, \bibinfo{journal}{Europhys. Lett.}
  \textbf{\bibinfo{volume}{89}}, \bibinfo{pages}{67005} (\bibinfo{year}{2010}).

\bibitem[{\citenamefont{Buitelaar et~al.}(2008)\citenamefont{Buitelaar,
  Fransson, Cantone, Smith, Anderson, Ardavan, Khlobystov, Morley, Porfyrakis,
  and Briggs}}]{Buitelaar2008}
\bibinfo{author}{\bibfnamefont{M.~R.} \bibnamefont{Buitelaar}},
  \bibinfo{author}{\bibfnamefont{J.}~\bibnamefont{Fransson}},
  \bibinfo{author}{\bibfnamefont{A.~L.} \bibnamefont{Cantone}},
  \bibinfo{author}{\bibfnamefont{C.~G.} \bibnamefont{Smith}},
  \bibinfo{author}{\bibfnamefont{D.}~\bibnamefont{Anderson}},
  \bibinfo{author}{\bibfnamefont{A.}~\bibnamefont{Ardavan}},
  \bibinfo{author}{\bibfnamefont{A.~N.} \bibnamefont{Khlobystov}},
  \bibinfo{author}{\bibfnamefont{G.~W.} \bibnamefont{Morley}},
  \bibinfo{author}{\bibfnamefont{K.}~\bibnamefont{Porfyrakis}},
  \bibnamefont{and} \bibinfo{author}{\bibfnamefont{G.~A.~D.}
  \bibnamefont{Briggs}}, \bibinfo{journal}{Phys. Rev. B}
  \textbf{\bibinfo{volume}{77}}, \bibinfo{pages}{245439}
  (\bibinfo{year}{2008}).

\bibitem[{\citenamefont{Ponomarenko et~al.}(2008)\citenamefont{Ponomarenko,
  Schedin, Katsnelson, Yang, Hill, Novoselov, and Geim}}]{Ponomarenko2008}
\bibinfo{author}{\bibfnamefont{L.~A.} \bibnamefont{Ponomarenko}},
  \bibinfo{author}{\bibfnamefont{F.}~\bibnamefont{Schedin}},
  \bibinfo{author}{\bibfnamefont{M.~I.} \bibnamefont{Katsnelson}},
  \bibinfo{author}{\bibfnamefont{R.}~\bibnamefont{Yang}},
  \bibinfo{author}{\bibfnamefont{E.~W.} \bibnamefont{Hill}},
  \bibinfo{author}{\bibfnamefont{K.~S.} \bibnamefont{Novoselov}},
  \bibnamefont{and} \bibinfo{author}{\bibfnamefont{A.~K.} \bibnamefont{Geim}},
  \bibinfo{journal}{Science} \textbf{\bibinfo{volume}{320}},
  \bibinfo{pages}{356} (\bibinfo{year}{2008}).

\bibitem[{\citenamefont{Stampfer
  et~al.}(2009{\natexlab{a}})\citenamefont{Stampfer, Guttinger, Molitor, Graf,
  Ihn, and Ensslin}}]{Stampfer2008}
\bibinfo{author}{\bibfnamefont{C.}~\bibnamefont{Stampfer}},
  \bibinfo{author}{\bibfnamefont{J.}~\bibnamefont{Guttinger}},
  \bibinfo{author}{\bibfnamefont{F.}~\bibnamefont{Molitor}},
  \bibinfo{author}{\bibfnamefont{D.}~\bibnamefont{Graf}},
  \bibinfo{author}{\bibfnamefont{T.}~\bibnamefont{Ihn}}, \bibnamefont{and}
  \bibinfo{author}{\bibfnamefont{K.}~\bibnamefont{Ensslin}},
  \bibinfo{journal}{Appl. Phys. Lett.} \textbf{\bibinfo{volume}{92}},
  \bibinfo{pages}{012102} (\bibinfo{year}{2009}{\natexlab{a}}).

\bibitem[{\citenamefont{Dorn et~al.}(2004{\natexlab{a}})\citenamefont{Dorn,
  Ihn, Ensslin, Wegscheider, and Bichler}}]{Schnez2010}
\bibinfo{author}{\bibfnamefont{A.}~\bibnamefont{Dorn}},
  \bibinfo{author}{\bibfnamefont{T.}~\bibnamefont{Ihn}},
  \bibinfo{author}{\bibfnamefont{K.}~\bibnamefont{Ensslin}},
  \bibinfo{author}{\bibfnamefont{W.}~\bibnamefont{Wegscheider}},
  \bibnamefont{and} \bibinfo{author}{\bibfnamefont{M.}~\bibnamefont{Bichler}},
  \bibinfo{journal}{Phys. Rev. B} \textbf{\bibinfo{volume}{70}},
  \bibinfo{pages}{205306} (\bibinfo{year}{2004}{\natexlab{a}}).

\bibitem[{\citenamefont{Stampfer
  et~al.}(2009{\natexlab{b}})\citenamefont{Stampfer, Guttinger, Hellmuller,
  Molitor, Ensslin, , and Ihn}}]{Stampfer2009}
\bibinfo{author}{\bibfnamefont{C.}~\bibnamefont{Stampfer}},
  \bibinfo{author}{\bibfnamefont{J.}~\bibnamefont{Guttinger}},
  \bibinfo{author}{\bibfnamefont{S.}~\bibnamefont{Hellmuller}},
  \bibinfo{author}{\bibfnamefont{F.}~\bibnamefont{Molitor}},
  \bibinfo{author}{\bibfnamefont{K.}~\bibnamefont{Ensslin}}, ,
  \bibnamefont{and} \bibinfo{author}{\bibfnamefont{T.}~\bibnamefont{Ihn}},
  \bibinfo{journal}{Phys. Rev. Lett.} \textbf{\bibinfo{volume}{102}},
  \bibinfo{pages}{056403} (\bibinfo{year}{2009}{\natexlab{b}}).

\bibitem[{\citenamefont{Todd et~al.}(2009)\citenamefont{Todd, Chou, Amasha, and
  Goldhaber-Gordon}}]{Todd2009}
\bibinfo{author}{\bibfnamefont{K.}~\bibnamefont{Todd}},
  \bibinfo{author}{\bibfnamefont{H.-T.} \bibnamefont{Chou}},
  \bibinfo{author}{\bibfnamefont{S.}~\bibnamefont{Amasha}}, \bibnamefont{and}
  \bibinfo{author}{\bibfnamefont{D.}~\bibnamefont{Goldhaber-Gordon}},
  \bibinfo{journal}{Nano Letters} \textbf{\bibinfo{volume}{9}},
  \bibinfo{pages}{416} (\bibinfo{year}{2009}).

\bibitem[{\citenamefont{Chen et~al.}(2007)\citenamefont{Chen, Lin, Rooks, and
  Avouris}}]{Chen2007}
\bibinfo{author}{\bibfnamefont{Z.}~\bibnamefont{Chen}},
  \bibinfo{author}{\bibfnamefont{Y.~M.} \bibnamefont{Lin}},
  \bibinfo{author}{\bibfnamefont{.~J.} \bibnamefont{Rooks}}, \bibnamefont{and}
  \bibinfo{author}{\bibfnamefont{P.}~\bibnamefont{Avouris}},
  \bibinfo{journal}{Physica E} \textbf{\bibinfo{volume}{40}},
  \bibinfo{pages}{228} (\bibinfo{year}{2007}).

\bibitem[{\citenamefont{Han et~al.}(2007)\citenamefont{Han, Ozyilmaz, Zhang,
  and Kim}}]{Han2007}
\bibinfo{author}{\bibfnamefont{M.~Y.} \bibnamefont{Han}},
  \bibinfo{author}{\bibfnamefont{B.}~\bibnamefont{Ozyilmaz}},
  \bibinfo{author}{\bibfnamefont{Y.}~\bibnamefont{Zhang}}, \bibnamefont{and}
  \bibinfo{author}{\bibfnamefont{P.}~\bibnamefont{Kim}},
  \bibinfo{journal}{Phys. Rev. Lett} \textbf{\bibinfo{volume}{98}},
  \bibinfo{pages}{206805} (\bibinfo{year}{2007}).

\bibitem[{\citenamefont{Molitor et~al.}(2009)\citenamefont{Molitor, Jacobsen,
  Stampfer, Guttinger, Ihn, and Ensslin}}]{Molitor2009}
\bibinfo{author}{\bibfnamefont{F.}~\bibnamefont{Molitor}},
  \bibinfo{author}{\bibfnamefont{A.}~\bibnamefont{Jacobsen}},
  \bibinfo{author}{\bibfnamefont{C.}~\bibnamefont{Stampfer}},
  \bibinfo{author}{\bibfnamefont{J.}~\bibnamefont{Guttinger}},
  \bibinfo{author}{\bibfnamefont{T.}~\bibnamefont{Ihn}}, \bibnamefont{and}
  \bibinfo{author}{\bibfnamefont{K.}~\bibnamefont{Ensslin}},
  \bibinfo{journal}{Phys. Rev. B} \textbf{\bibinfo{volume}{79}},
  \bibinfo{pages}{075426} (\bibinfo{year}{2009}).

\bibitem[{\citenamefont{Liu et~al.}(2009)\citenamefont{Liu, Oostinga, Morpurgo,
  and Vandersypen}}]{Liu2009}
\bibinfo{author}{\bibfnamefont{X.}~\bibnamefont{Liu}},
  \bibinfo{author}{\bibfnamefont{J.~B.} \bibnamefont{Oostinga}},
  \bibinfo{author}{\bibfnamefont{A.~F.} \bibnamefont{Morpurgo}},
  \bibnamefont{and} \bibinfo{author}{\bibfnamefont{L.~M.~K.}
  \bibnamefont{Vandersypen}}, \bibinfo{journal}{Phys. Rev. B}
  \textbf{\bibinfo{volume}{80}}, \bibinfo{pages}{121407}
  (\bibinfo{year}{2009}).

\bibitem[{\citenamefont{Han et~al.}(2010)\citenamefont{Han, Brant, and
  Kim}}]{Han2010}
\bibinfo{author}{\bibfnamefont{M.~Y.} \bibnamefont{Han}},
  \bibinfo{author}{\bibfnamefont{J.~C.} \bibnamefont{Brant}}, \bibnamefont{and}
  \bibinfo{author}{\bibfnamefont{P.}~\bibnamefont{Kim}},
  \bibinfo{journal}{Phys. Rev. Lett} \textbf{\bibinfo{volume}{104}},
  \bibinfo{pages}{056801} (\bibinfo{year}{2010}).

\bibitem[{\citenamefont{Evaldsson et~al.}(2008)\citenamefont{Evaldsson,
  Zozoulenko, Xu, and Heinzel}}]{Evaldsson2008}
\bibinfo{author}{\bibfnamefont{M.}~\bibnamefont{Evaldsson}},
  \bibinfo{author}{\bibfnamefont{I.~V.} \bibnamefont{Zozoulenko}},
  \bibinfo{author}{\bibfnamefont{H.}~\bibnamefont{Xu}}, \bibnamefont{and}
  \bibinfo{author}{\bibfnamefont{T.}~\bibnamefont{Heinzel}},
  \bibinfo{journal}{Phys. Rev. B} \textbf{\bibinfo{volume}{78}},
  \bibinfo{pages}{161407} (\bibinfo{year}{2008}).

\bibitem[{\citenamefont{Adam et~al.}(2008)\citenamefont{Adam, Cho, Fuhrer, and
  Sarma}}]{Adam2008}
\bibinfo{author}{\bibfnamefont{S.}~\bibnamefont{Adam}},
  \bibinfo{author}{\bibfnamefont{S.}~\bibnamefont{Cho}},
  \bibinfo{author}{\bibfnamefont{M.~S.} \bibnamefont{Fuhrer}},
  \bibnamefont{and} \bibinfo{author}{\bibfnamefont{S.~D.} \bibnamefont{Sarma}},
  \bibinfo{journal}{Phys. Rev. Lett.} \textbf{\bibinfo{volume}{101}},
  \bibinfo{pages}{046404} (\bibinfo{year}{2008}).

\bibitem[{\citenamefont{Sols et~al.}(2007)\citenamefont{Sols, Guinea, and
  Neto}}]{Sols2007}
\bibinfo{author}{\bibfnamefont{F.}~\bibnamefont{Sols}},
  \bibinfo{author}{\bibfnamefont{F.}~\bibnamefont{Guinea}}, \bibnamefont{and}
  \bibinfo{author}{\bibfnamefont{A.~H.~C.} \bibnamefont{Neto}},
  \bibinfo{journal}{Phys. Rev. Lett.} \textbf{\bibinfo{volume}{99}},
  \bibinfo{pages}{166803} (\bibinfo{year}{2007}).

\bibitem[{\citenamefont{Casiraghi et~al.}(2007)\citenamefont{Casiraghi,
  Hartschuh, Lidorikis, Qian, Harutyunyan, Gokus, Novoselov, and
  Ferrari}}]{CasiraghiNL}
\bibinfo{author}{\bibfnamefont{C.}~\bibnamefont{Casiraghi}},
  \bibinfo{author}{\bibfnamefont{A.}~\bibnamefont{Hartschuh}},
  \bibinfo{author}{\bibfnamefont{E.}~\bibnamefont{Lidorikis}},
  \bibinfo{author}{\bibfnamefont{H.}~\bibnamefont{Qian}},
  \bibinfo{author}{\bibfnamefont{H.}~\bibnamefont{Harutyunyan}},
  \bibinfo{author}{\bibfnamefont{T.}~\bibnamefont{Gokus}},
  \bibinfo{author}{\bibfnamefont{K.~S.} \bibnamefont{Novoselov}},
  \bibnamefont{and} \bibinfo{author}{\bibfnamefont{A.~C.}
  \bibnamefont{Ferrari}}, \bibinfo{journal}{Nano Lett.}
  \textbf{\bibinfo{volume}{7}}, \bibinfo{pages}{2711} (\bibinfo{year}{2007}).

\bibitem[{\citenamefont{Ferrari et~al.}(2006)\citenamefont{Ferrari, Meyer,
  Scardaci, Casiraghi, Lazzeri, Mauri, Piscanec, Jiang, Novoselov, Roth
  et~al.}}]{Ferrari2006}
\bibinfo{author}{\bibfnamefont{A.~C.} \bibnamefont{Ferrari}},
  \bibinfo{author}{\bibfnamefont{J.~C.} \bibnamefont{Meyer}},
  \bibinfo{author}{\bibfnamefont{V.}~\bibnamefont{Scardaci}},
  \bibinfo{author}{\bibfnamefont{C.}~\bibnamefont{Casiraghi}},
  \bibinfo{author}{\bibfnamefont{M.}~\bibnamefont{Lazzeri}},
  \bibinfo{author}{\bibfnamefont{F.}~\bibnamefont{Mauri}},
  \bibinfo{author}{\bibfnamefont{S.}~\bibnamefont{Piscanec}},
  \bibinfo{author}{\bibfnamefont{D.}~\bibnamefont{Jiang}},
  \bibinfo{author}{\bibfnamefont{K.}~\bibnamefont{Novoselov}},
  \bibinfo{author}{\bibfnamefont{S.}~\bibnamefont{Roth}}, \bibnamefont{et~al.},
  \bibinfo{journal}{Phys. Rev. Lett} \textbf{\bibinfo{volume}{97}},
  \bibinfo{pages}{187401} (\bibinfo{year}{2006}).

\bibitem[{\citenamefont{Horcas et~al.}(2007)\citenamefont{Horcas, Fernandez,
  Gomez-Rodriguez, Colchero, Gomez-Herrero, and Baro}}]{Horcas2007}
\bibinfo{author}{\bibfnamefont{I.}~\bibnamefont{Horcas}},
  \bibinfo{author}{\bibfnamefont{R.}~\bibnamefont{Fernandez}},
  \bibinfo{author}{\bibfnamefont{J.~M.} \bibnamefont{Gomez-Rodriguez}},
  \bibinfo{author}{\bibfnamefont{J.}~\bibnamefont{Colchero}},
  \bibinfo{author}{\bibfnamefont{J.}~\bibnamefont{Gomez-Herrero}},
  \bibnamefont{and} \bibinfo{author}{\bibfnamefont{A.~M.} \bibnamefont{Baro}},
  \bibinfo{journal}{Rev. Sci. Instrum.} \textbf{\bibinfo{volume}{78}},
  \bibinfo{pages}{013705} (\bibinfo{year}{2007}).

\bibitem[{\citenamefont{Woodside and McEuen}(2002)}]{Woodside2002}
\bibinfo{author}{\bibfnamefont{M.~T.} \bibnamefont{Woodside}} \bibnamefont{and}
  \bibinfo{author}{\bibfnamefont{P.~L.} \bibnamefont{McEuen}},
  \bibinfo{journal}{Science} \textbf{\bibinfo{volume}{296}},
  \bibinfo{pages}{1098} (\bibinfo{year}{2002}).

\bibitem[{\citenamefont{Pioda et~al.}(2004)\citenamefont{Pioda, Kicin, Ihn,
  Sigrist, Fuhrer, Ensslin, A.Weichselbaum, Ulloa, Reinwald, and
  Wegscheider}}]{Pioda2004}
\bibinfo{author}{\bibfnamefont{A.}~\bibnamefont{Pioda}},
  \bibinfo{author}{\bibfnamefont{S.}~\bibnamefont{Kicin}},
  \bibinfo{author}{\bibfnamefont{T.}~\bibnamefont{Ihn}},
  \bibinfo{author}{\bibfnamefont{M.}~\bibnamefont{Sigrist}},
  \bibinfo{author}{\bibfnamefont{A.}~\bibnamefont{Fuhrer}},
  \bibinfo{author}{\bibfnamefont{K.}~\bibnamefont{Ensslin}},
  \bibinfo{author}{\bibnamefont{A.Weichselbaum}},
  \bibinfo{author}{\bibfnamefont{S.~E.} \bibnamefont{Ulloa}},
  \bibinfo{author}{\bibfnamefont{M.}~\bibnamefont{Reinwald}}, \bibnamefont{and}
  \bibinfo{author}{\bibfnamefont{W.}~\bibnamefont{Wegscheider}},
  \bibinfo{journal}{Phys. Rev. Lett.} \textbf{\bibinfo{volume}{93}},
  \bibinfo{pages}{216801} (\bibinfo{year}{2004}).

\bibitem[{\citenamefont{Connolly et~al.}(2010)\citenamefont{Connolly, Chiou,
  Smith, Anderson, Jones, Lombardo, Fasoli, and Ferrari}}]{Connolly2010}
\bibinfo{author}{\bibfnamefont{M.~R.} \bibnamefont{Connolly}},
  \bibinfo{author}{\bibfnamefont{K.~L.} \bibnamefont{Chiou}},
  \bibinfo{author}{\bibfnamefont{C.~G.} \bibnamefont{Smith}},
  \bibinfo{author}{\bibfnamefont{D.}~\bibnamefont{Anderson}},
  \bibinfo{author}{\bibfnamefont{G.~A.~C.} \bibnamefont{Jones}},
  \bibinfo{author}{\bibfnamefont{A.}~\bibnamefont{Lombardo}},
  \bibinfo{author}{\bibfnamefont{A.}~\bibnamefont{Fasoli}}, \bibnamefont{and}
  \bibinfo{author}{\bibfnamefont{A.~C.} \bibnamefont{Ferrari}},
  \bibinfo{journal}{Appl. Phys. Lett.} \textbf{\bibinfo{volume}{96}},
  \bibinfo{pages}{113501} (\bibinfo{year}{2010}).

\bibitem[{\citenamefont{Ni et~al.}(2009)\citenamefont{Ni, Yu, Qiang, Wang, Liu,
  Wong, Miao, Huang, and Shen}}]{Ni2009}
\bibinfo{author}{\bibfnamefont{Z.~H.} \bibnamefont{Ni}},
  \bibinfo{author}{\bibfnamefont{T.}~\bibnamefont{Yu}},
  \bibinfo{author}{\bibfnamefont{Z.}~\bibnamefont{Qiang}},
  \bibinfo{author}{\bibfnamefont{Y.~Y.} \bibnamefont{Wang}},
  \bibinfo{author}{\bibfnamefont{L.}~\bibnamefont{Liu}},
  \bibinfo{author}{\bibfnamefont{C.~P.} \bibnamefont{Wong}},
  \bibinfo{author}{\bibfnamefont{J.}~\bibnamefont{Miao}},
  \bibinfo{author}{\bibfnamefont{W.}~\bibnamefont{Huang}}, \bibnamefont{and}
  \bibinfo{author}{\bibfnamefont{Z.~X.} \bibnamefont{Shen}},
  \bibinfo{journal}{ACS Nano} \textbf{\bibinfo{volume}{3}},
  \bibinfo{pages}{569} (\bibinfo{year}{2009}).

\bibitem[{\citenamefont{Kicin et~al.}(2005)\citenamefont{Kicin, Pioda, Ihn,
  Sigrist, Fuhrer, Ensslin, Reinwald, and Wegscheider}}]{Kicin2005}
\bibinfo{author}{\bibfnamefont{S.}~\bibnamefont{Kicin}},
  \bibinfo{author}{\bibfnamefont{A.}~\bibnamefont{Pioda}},
  \bibinfo{author}{\bibfnamefont{T.}~\bibnamefont{Ihn}},
  \bibinfo{author}{\bibfnamefont{M.}~\bibnamefont{Sigrist}},
  \bibinfo{author}{\bibfnamefont{A.}~\bibnamefont{Fuhrer}},
  \bibinfo{author}{\bibfnamefont{K.}~\bibnamefont{Ensslin}},
  \bibinfo{author}{\bibfnamefont{M.}~\bibnamefont{Reinwald}}, \bibnamefont{and}
  \bibinfo{author}{\bibfnamefont{W.}~\bibnamefont{Wegscheider}},
  \bibinfo{journal}{New Journal of Physics} \textbf{\bibinfo{volume}{185}},
  \bibinfo{pages}{1367} (\bibinfo{year}{2005}).

\bibitem[{\citenamefont{Dorn et~al.}(2004{\natexlab{b}})\citenamefont{Dorn,
  Ihn, Ensslin, Wegscheider, and Bichler}}]{Dorn2004}
\bibinfo{author}{\bibfnamefont{A.}~\bibnamefont{Dorn}},
  \bibinfo{author}{\bibfnamefont{T.}~\bibnamefont{Ihn}},
  \bibinfo{author}{\bibfnamefont{K.}~\bibnamefont{Ensslin}},
  \bibinfo{author}{\bibfnamefont{W.}~\bibnamefont{Wegscheider}},
  \bibnamefont{and} \bibinfo{author}{\bibfnamefont{M.}~\bibnamefont{Bichler}},
  \bibinfo{journal}{Phys. Rev. B} \textbf{\bibinfo{volume}{70}},
  \bibinfo{pages}{205306} (\bibinfo{year}{2004}{\natexlab{b}}).

\bibitem[{\citenamefont{Bieszynski et~al.}(2007)\citenamefont{Bieszynski,
  Zwanenburg, Westerwelt, Roest, Bakkers, and Kouwenhoven}}]{Bieszynski2007}
\bibinfo{author}{\bibfnamefont{A.~C.} \bibnamefont{Bieszynski}},
  \bibinfo{author}{\bibfnamefont{F.~A.} \bibnamefont{Zwanenburg}},
  \bibinfo{author}{\bibfnamefont{R.~M.~.} \bibnamefont{Westerwelt}},
  \bibinfo{author}{\bibfnamefont{A.~L.} \bibnamefont{Roest}},
  \bibinfo{author}{\bibfnamefont{E.~P. A.~M.} \bibnamefont{Bakkers}},
  \bibnamefont{and} \bibinfo{author}{\bibfnamefont{L.~P.}
  \bibnamefont{Kouwenhoven}}, \bibinfo{journal}{Nano Letters}
  \textbf{\bibinfo{volume}{7}}, \bibinfo{pages}{2559} (\bibinfo{year}{2007}).

\bibitem[{\citenamefont{Gildemeister et~al.}(2007)\citenamefont{Gildemeister,
  Ihn, Sigrist, Ensslin, Driscoll, and Gossard}}]{Gildemeister2007}
\bibinfo{author}{\bibfnamefont{A.~E.} \bibnamefont{Gildemeister}},
  \bibinfo{author}{\bibfnamefont{T.}~\bibnamefont{Ihn}},
  \bibinfo{author}{\bibfnamefont{M.}~\bibnamefont{Sigrist}},
  \bibinfo{author}{\bibfnamefont{K.}~\bibnamefont{Ensslin}},
  \bibinfo{author}{\bibfnamefont{D.~C.} \bibnamefont{Driscoll}},
  \bibnamefont{and} \bibinfo{author}{\bibfnamefont{A.~C.}
  \bibnamefont{Gossard}}, \bibinfo{journal}{Phys. Rev. B}
  \textbf{\bibinfo{volume}{75}}, \bibinfo{pages}{195338}
  (\bibinfo{year}{2007}).

\bibitem[{\citenamefont{Wilson and Cobden}(2008)}]{Wilson2008}
\bibinfo{author}{\bibfnamefont{N.~R.} \bibnamefont{Wilson}} \bibnamefont{and}
  \bibinfo{author}{\bibfnamefont{D.~H.} \bibnamefont{Cobden}},
  \bibinfo{journal}{Nano Letters} \textbf{\bibinfo{volume}{8}},
  \bibinfo{pages}{2161} (\bibinfo{year}{2008}).

\bibitem[{\citenamefont{Fallahi et~al.}(2005)\citenamefont{Fallahi, Bleszynski,
  Westervelt, Huang, Walls, Heller, Hanson, and Gossard}}]{Fallahi2005}
\bibinfo{author}{\bibfnamefont{P.}~\bibnamefont{Fallahi}},
  \bibinfo{author}{\bibfnamefont{A.~C.} \bibnamefont{Bleszynski}},
  \bibinfo{author}{\bibfnamefont{R.~M.} \bibnamefont{Westervelt}},
  \bibinfo{author}{\bibfnamefont{J.}~\bibnamefont{Huang}},
  \bibinfo{author}{\bibfnamefont{J.~D.} \bibnamefont{Walls}},
  \bibinfo{author}{\bibfnamefont{E.~J.} \bibnamefont{Heller}},
  \bibinfo{author}{\bibfnamefont{M.}~\bibnamefont{Hanson}}, \bibnamefont{and}
  \bibinfo{author}{\bibfnamefont{A.~C.} \bibnamefont{Gossard}},
  \bibinfo{journal}{Nano Letters} \textbf{\bibinfo{volume}{5}},
  \bibinfo{pages}{223} (\bibinfo{year}{2005}).

\bibitem[{\citenamefont{Ruzin et~al.}(1992)\citenamefont{Ruzin, Chandrasekhar,
  Levin, and Glazman}}]{Ruzin1992}
\bibinfo{author}{\bibfnamefont{I.~M.} \bibnamefont{Ruzin}},
  \bibinfo{author}{\bibfnamefont{V.}~\bibnamefont{Chandrasekhar}},
  \bibinfo{author}{\bibfnamefont{E.~I.} \bibnamefont{Levin}}, \bibnamefont{and}
  \bibinfo{author}{\bibfnamefont{L.~I.} \bibnamefont{Glazman}},
  \bibinfo{journal}{Phys. Rev. B} \textbf{\bibinfo{volume}{45}},
  \bibinfo{pages}{13469} (\bibinfo{year}{1992}).

\bibitem[{\citenamefont{Gallagher et~al.}(2010)\citenamefont{Gallagher, Todd,
  and Goldhaber-Gordon}}]{Gallagher2010}
\bibinfo{author}{\bibfnamefont{P.}~\bibnamefont{Gallagher}},
  \bibinfo{author}{\bibfnamefont{K.}~\bibnamefont{Todd}}, \bibnamefont{and}
  \bibinfo{author}{\bibfnamefont{D.}~\bibnamefont{Goldhaber-Gordon}},
  \bibinfo{journal}{Phys. Rev. B} \textbf{\bibinfo{volume}{81}},
  \bibinfo{pages}{115409} (\bibinfo{year}{2010}).

\end{thebibliography}

\end{document}